\documentclass[11pt,a4paper]{article}
\usepackage{jcappub}
\usepackage{mathrsfs}
\usepackage{epsfig}
\usepackage{cite}
\usepackage{amssymb}
\usepackage{rotating}
\usepackage{amsmath}
\usepackage{amssymb}
\usepackage{graphicx}
\usepackage{color}
\usepackage{amsmath}
\usepackage{amsthm}
\usepackage{slashed}
\usepackage{soul}
\usepackage[usenames,dvipsnames,svgnames,table]{xcolor}

\title{Cosmology in a Globally U(1) Symmetric Scalar-Tensor Gravity}

\author[1]{Meir Shimon,}\emailAdd{meirs@wise.tau.ac.il}
\affiliation[1]{School of Physics and Astronomy, Tel Aviv University, Tel Aviv
69978, Israel}

\author{}

\abstract{A cosmological model is formulated 
in the context of a scalar-tensor theory of gravity in which the entire cosmic 
background evolution is due to a complex scalar field evolving in Minkowski 
spacetime, such that its (dimensional) modulus 
is conformally coupled, and the (dimensionless) phase 
is only minimally coupled to gravitation. 
The former regulates the dynamics of masses; cosmological redshift reflects 
the growth of particle masses over cosmological time scales, 
not space expansion.
An interplay between the energy density of radiation 
and that of the kinetic energy associated with the phase 
(which are of opposite relative signs) 
results in a non-singular cosmological model that encompases 
the observed redshifting phase preceded by a turnaround 
that follows a blushifting phase. 
The model is essentially free of any horizon, flatness or anisotropy problems.
Quantum excitations of the phase during the 
matter dominated blueshifting era generate a flat spectrum of adiabatic gaussian 
scalar perturbations on cosmological 
scales. No detectable primordial tensor modes 
are generated in this scenario, and cold dark matter must be 
fermionic. Other consequences are also discussed.}


\begin{document}

\maketitle

\flushbottom

\section{Introduction}
The standard cosmological model with an early inflationary scenario has 
clearly been a very successful paradigm that provides a compelling interpretation 
of essentially all current cosmic microwave background 
(CMB), large scale structure (LSS) measurements, and the agreement between Big Bang 
nucleosynthesis (BBN) predictions and light element abundances.
It is remarkable that the cosmological model provides a very good fit to 
extensive observational data, that sample phenomena over a vast dynamical range, using 
less than a dozen free parameters. 

However, the essence of dark energy (DE) and cold dark matter (CDM) 
-- two key ingredients in the model that determine the background evolution, 
LSS formation history, and gravitational potential on galactic 
scales -- remain elusive. Additionally, 
what is considered by many as the most pristine fingerprint of cosmic inflation 
[1-3] -- a major underpinning of the standard cosmological 
model -- B-mode polarization of the CMB [4-6] induced by primordial 
gravitational waves (PGW), has not been detected. 
The latter is admittedly a very challenging measurement in the 
presence of e.g., polarized Galactic dust, nonlinear density perturbations, and 
instrumental systematics. In light of these hurdles and 
the theoretically viable broad window for 
the energy scale of inflation, it is not unlikely that this signal will never be 
measured at sufficient statistical significance. Yet, its non-detection will not 
rule out the inflationary paradigm, but would rather set an (arguably very weak) 
upper bound on the energy scale of inflation. 

According to the standard cosmological model global 
evolution is driven by space expansion, namely the time-dependent Hubble scale provides 
the `clock' for the evolving properties 
of radiation and matter, resulting in a sequence of cosmological epochs. This clock is only 
meaningful if other time scales, e.g., the Planck time, or characteristic Compton times, 
evolve differently, particularly if they are 
non-varying; thus, space expansion is a relative notion.

The main objective of the present work is to demonstrate the viability of 
an alternative, non-singular `bouncing' cosmological model within a 
physical framework based on a $U(1)$ symmetric scalar-tensor theory of gravity.
Spacetime, effectively Minkowski, is static in this model.
Other non-scalar fields, e.g. Dirac, Weyl, and electromagnetic fields 
are similarly non-evolving on this static background space.
Cosmic evolution is then achieved by evolving scalar fields, i.e. 
evolution of fundamental particle masses and the Planck mass, 
while their dimensionless ratios remain fixed to their standard values. 
The time-dependent scalar field which regulates the (dynamical) masses 
of particles starts infinitely large, monotonically decreases until it `bounces', 
then grows again unboundedly. In other words, the Compton wavelength 
associated with fundamental particles starts 
infinitely small, increases until it peaks at the `bounce', 
then decreases again. 
Described in terms of these length `units' the universe is 
said to undergo a `contraction' epoch, 
followed by a `bounce' and `expansion'.

Since the universe does not actually 
contract and expand in the proposed model, it is perhaps 
more appropriate to refer to these epochs in a frame-independent 
fashion that is faithful to what is actually observed as 
blue-and red-shifting 
eras, respectively. For similar reasons it is more appropriate to 
refer to the `bounce' by `turnaround' or `turnaround point' 
between the blue-and red-shifting eras. For the rest of this work 
we use these terms in reference to the proposed model 
instead of the more traditional ones.

We explore a wide range of the possible ramifications of the model 
(albeit not exhaustively) which {\it a priori} demotes the gravitational 
constant, particle masses, and all other dimensional constants, from 
their fundamental-physical-constants status, and replaces them with 
the conformally-coupled modulus of a single complex field.

A dynamical vacuum expectation value (VEV) for the Higgs field 
is naturally accommodated by a Weyl-symmetric 
theory that essentially incorporates all the fundamental interactions into 
this framework [7]. 
In particular, the standard model (SM) of particle physics, where the Higgs VEV 
is a fixed constant, is just one convenient gauge choice. Applying a 
Weyl transformation to such field configurations that appropriately endows 
the Higgs VEV with exactly the same dynamics of the evolving scalar field 
in the background cosmological model guarantees that the dynamical particle 
masses are continuous everywhere, exactly as in the SM 
and general relativity (GR). Thus, a static field configuration, e.g. a planet or a galaxy, 
transforms into a stationary one, while observables in a planet or a star 
are the same as in standard physics.

While the intriguing possibility that the cosmological redshift could be explained 
by means of time-dependent fundamental `constants' is nearly as old as 
(what has become) the standard expanding space interpretation 
[8], it has been waived off by big bang proponents as soon as it was proposed [9]. 
This basic idea has re-emerged later within the framework of e.g., 
scalar-tensor theories of gravity [10-14] and in the context of Weyl-geometry, e.g. [15]. 

Temporal variation of the gravitational constant, $G$, is a key feature in 
scalar-tensor theories, of which Brans-Dicke (BD) theory is archetypical. 
Standard interpretation of observational constraints, e.g. [16], 
usually renders this theory equivalent to GR due mainly to the convention 
that all other (dimensional) fundamental quantities are constant. 
Our approach is fundamentally 
different as it is based on the premise that 
{\it all} fundamental length scales have exactly the same dynamics 
which is regulated by the modulus of the complex scalar field. 
In other words, the often cited $\omega_{BD}>40000$ [16] that essentially 
fixes the BD scalar field (i.e. Newton's $G$) to a constant thereby reducing 
the theory to GR, 
does not apply to the Bergmann-Wagoner type of theory 
[17, 18]
studied here. In particular, our approach guarantees that 
{\it dimensionless} observables (in a sense that will 
be more clearly defined below) 
are by construction unchanged compared to their corresponding values in 
the standard cosmological model in the domain of the latter validity, 
i.e. down to the `bounce'. 

Throughout, we adopt a mostly-positive signature for the 
spacetime metric $(-1,1,1,1)$. Our units convention is $\hbar=c=k_{B}=1$.
We outline our theoretical approach in section 2, 
and the cosmological model is presented 
in section 3. In Section 4 we discuss and summarize our main results. 

\section{Theoretical Framework}

Consider the following scalar-tensor theory of the 
Bergmann-Wagoner type [17, 18]
\begin{eqnarray}
\mathcal{I}&=&\int\left[\frac{1}{2}F(\phi^{K})R
-\frac{1}{2}\mathcal{G}_{IJ}(\phi^{K})\phi_{\mu}^{I}\phi^{J\mu}-V(\phi^{K})
+\mathcal{L}_{M}(\phi^{K})\right]\sqrt{-g}d^{4}x, 
\end{eqnarray}
where $\mathcal{G}_{IJ}$ is a two-dimensional metric in field space 
(that possibly depends on the fields), and capital Latin letters assume the 
values 1 \& 2. A useful basis is $\phi^{(1)}=\rho e^{i\theta}$ \&  $\phi^{(2)}=(\phi^{(1)})^{*}$. 
The only non-vanishing components of $\mathcal{G}_{IJ}$ in this basis are the {\it constant} 
off-diagonal $\mathcal{G}_{12}=\mathcal{G}_{21}=-1$ components. In this complex fields frame $F\equiv\frac{|\phi|^{2}}{3}$ 
\& $V=\lambda|\phi|^{4}$ where here $|\phi|\equiv|\phi^{(1)}|=|\phi^{(2)}|$.
Alternatively, in a polar field frame with 
real $\phi^{(1)}=\rho$ \&  $\phi^{(2)}=\theta$ (which correspond to the modulus and phase, 
respectively, of the field $\phi^{(1)}$ of the complex field frame),
the fields space metric is $\mathcal{G}_{IJ}=-2\cdot diag(1,\rho^{2})$, 
and $F\equiv\frac{\rho^{2}}{3}$ \& $V=\lambda\rho^{4}$.
In sections 2 \& 3 we employ mainly the complex and real 
fields frames, respectively. 
Eq. (2.1) differs from the BD action in that $\rho$ appears not only in the 
curvature and kinetic terms but also in $\mathcal{L}_{M}$ and $V$.
Eq. (2.1) is Weyl-symmetric in a sense that will be made clear below.
Here and throughout, Greek indices run over spacetime coordinates, 
and $f_{\mu}\equiv\frac{\partial f}{\partial x^{\mu}}$ 
for any function $f$.

Employing $\mathcal{G}_{IJ}=diag(-2,-2\rho^{2})$ 
[or even $\mathcal{G}_{IJ}=diag(-2,2\rho^{2})$ for that matter] 
guarantees that gravitation is attractive rather 
than repulsive provided that the energy density 
$\rho_{M}$ satisfies $\rho_{M}=-\mathcal{L}_{M}$ 
in the non-relativistic (NR) limit. 
Since $\mathcal{L}_{M}$, $V$ \& $F$ depend only on $\rho$ 
then $\theta$ is a free field that only 
minimally couples to gravity  
in this theory with a kinetic term of the canonical sign. 
The reason for it being only minimally coupled to gravity 
is that it is dimensionless and therefore blind to Weyl transformations. 
The fact that $\theta$ is massless simply reflects the assumed $U(1)$ 
symmetry. In case that this symmetry is only approximate then $\theta$ is 
in general massive. For the rest of this work we assume that $U(1)$ 
is an exact {\it global} symmetry of Eq. (2.1).

A necessary but insufficient requirement from a Weyl-symmetric 
action is that it contains no dimensional constants.
The specific pre-factor $\frac{F}{2}=\frac{\rho^{2}}{6}$ 
in the curvature term combined with the canonical kinetic term of the scalar 
field guarantee that the action described by Eq. (2.1) is 
invariant under the Weyl transformation (or rather {\it local} field redefinition) 
$g_{\mu\nu}\rightarrow\Omega^{2}g_{\mu\nu}$, $\rho\rightarrow\rho/\Omega$ 
(or equivalently $|\phi|\rightarrow|\phi|/\Omega$) 
and $\mathcal{L}_{M}\rightarrow\mathcal{L}_{M}/\Omega^{4}$ 
where $\Omega(x)$ is an arbitrary function, e.g. [19].  
Therefore, the proposed cosmological 
model is essentially described by a Weyl-symmetric theory 
-- or as put by a few what only looks as such, e.g. [20-23] -- through the entire 
cosmic history. Clearly, the single degree of 
freedom $\Omega$ is insufficient to simultaneously gauge-fix both $\rho$ \& $\theta$.
In particular, $\theta$ and its perturbations cannot be removed 
by Weyl transformations or by local gauge transformations.

Naively comparing the curvature term in Eq. (2.1) to the corresponding 
term in the Einstein-Hilbert (EH) action, while fixing $\phi$, implies the 
latter must be of order the Planck mass. However, 
$\mathcal{L}_{M}=\mathcal{L}_{M}(\rho)$ and therefore {\it all particle masses 
are proportional to $\rho$ and although it is dynamical, 
dimensionless ratios of particle masses 
and of particle masses to the Planck mass are 
fixed to their SM values}. This is a special property 
of the action described by Eq. (2.1) that significantly 
distinguishes it from other scalar-tensor theories in 
general, and BD in particular.

Variation of Eq. (2.1) with respect to $g_{\mu\nu}$ and $\phi^{I}$, 
results in the generalized Einstein equations, and scalar field equation, 
respectively, e.g. [24, 25]
\begin{eqnarray}
FG_{\mu}^{\nu}&=&-T_{M,\mu}^{\nu}
+\Theta_{\mu}^{\nu}-\delta_{\mu}^{\nu}V\\
\frac{F^{I}}{2}R+\Box\phi^{I}&+&\Gamma^{I}_{JK}\phi^{J;\mu}\phi_{;\mu}^{K}
-V^{I}+\mathcal{L}_{M}^{I}=0,
\end{eqnarray}
where $f^{I}\equiv\mathcal{G}^{IJ}\frac{\partial f}{\partial\phi^{J}}$ 
for any functional $f$. Here, $\Gamma^{I}_{JK}\equiv
\frac{1}{2}\mathcal{G}^{IL}(\mathcal{G}_{LJ,K}+\mathcal{G}_{LK,J}
-\mathcal{G}_{JK,L})$ is the `connection' 
computed from the two-dimensional field space metric $\mathcal{G}_{IJ}$.
Energy-momentum is not conserved,
\begin{eqnarray}
T_{M,\mu;\nu}^{\nu}&=&\mathcal{L}_{M,I}\phi_{\mu}^{I},
\end{eqnarray}
where semicolons stand for covariant derivatives.
The effective energy-momentum tensor $\Theta_{\mu}^{\nu}$ 
associated with the scalar fields is
\begin{eqnarray}
\Theta_{\mu}^{\nu}=\mathcal{G}_{IJ}(\phi_{\mu}^{I}\phi^{J,\nu}-\frac{1}{2}\delta_{\mu}^{\nu}
\phi^{I,\rho}\phi^{J}_{\rho})+F_{\mu}^{\nu}-\delta_{\mu}^{\nu}\Box F.
\end{eqnarray}
Here and throughout, $f_{\mu}^{\nu}\equiv(f_{,\mu})^{;\nu}$, 
the covariant Laplacian is $\Box f$, 
and $(T_{M})_{\mu\nu}\equiv\frac{2}{\sqrt{-g}}\frac{\delta(\sqrt{-g}\mathcal{L}_{M})}
{\delta g^{\mu\nu}}$ is the energy-momentum tensor. 
In the perfect fluid approximation the latter 
reads $(T_{M})_{\mu}^{\nu}\equiv\rho_{M}\cdot diag(-1,w,w,w)$ where $w$ 
is the equation of state (EOS) describing the fluid.
From the combination of Eq. (2.3) and the trace of Eq. (2.2) we obtain the constraint
\begin{eqnarray}
\phi^{I}\mathcal{L}_{M,I}=T_{M},
\end{eqnarray}
in both the polar and complex field frames defined below Eq. (2.1).
This constraint is indeed consistent with a traceless energy-momentum tensor 
but only in case that the matter Lagrangian density is independent of the 
scalar field, an often-made assumption (for example in GR or BD theory 
where all particle masses are fixed) that we relax in 
the present work. Therefore, traceless 
energy-momentum is a pre-requisite of Weyl-symmetric theories only 
when it is assumed that $\mathcal{L}_{M}$ does not depend on the scalar 
field.

It will be argued in section 3.1 that gravitation in the framework 
described by Eq. (2.1) is sourced only by the potential term of 
the $\phi$-dependent matter Lagrangian, while the kinetic term 
is in certain gauges the `curvature' term itself. 
Since $\mathcal{L}_{M}=-\rho_{M}$ it immediately 
follows from Eq. (2.6) that
\begin{eqnarray}
\rho_{M}\propto\rho^{1-3w}.
\end{eqnarray} 
Here, $\lambda\rho^{4}$ has been absorbed 
in $\mathcal{L}_{M}$ with an effective EOS parameter $w=-1$. 
As expected, $\rho_{M}$ is a quartic potential in the case $w=-1$, is independent of 
$\rho$, i.e. of masses, in the case $w=1/3$, and linear in masses in case of NR fermions, 
i.e. the $w=0$. For exactly the same reason CDM in the present model 
must be fermionic, 
in contrast with the standard cosmological model in which 
CDM could in principle be made up by either fermions or bosons. 
This, of course, is  
a direct result of Weyl invariance of the theory that does not allow 
any dimensional constants in Eq. (2.1). Therefore, CDM has to contribute 
a term $\propto\phi\bar{\psi}\psi$ to $\mathcal{L}_{M}$ in Eq. (2.1) where $\psi$ is 
some fermionic field.
In the case of `stiff' matter, $w_{stiff}=1$, 
$\rho_{M}\propto\rho^{-2}$. We assume that no contribution 
to the perfect fluid, described by $\mathcal{L}_{M}$, is characterized by 
a sound speed $c_{s}>1/\sqrt{3}$, i.e. $w>1/3$.

As was noted already, 
a solution $g_{\mu\nu}$ \& $\phi=(\rho,\theta)$ of the field equations (2.2) \& (2.3) 
is symmetric under Weyl (local) rescaling $g_{\mu\nu}\rightarrow\Omega^{2}g_{\mu\nu}$ 
\& $\rho\rightarrow\rho/\Omega$. 
Also as in GR, the metric `seen' by 
a massless test particle is $g_{\mu\nu}$ and the metric 
governing the kinematics of a massive particle 
with mass $m$ is $m^{2}g_{\mu\nu}$. 
But in contrast with GR, in which masses are assumed to be constant, 
in the present model $m\propto\rho$, which implies that the effective metric 
`seen' by a massive test particle is $\rho^{2}g_{\mu\nu}$. The latter is invariant 
under Weyl transformations, but $g_{\mu\nu}$ is not. However, null geodesics 
are blind to Weyl transformations, and the metric seen by 
massless particles is effectively $g_{\mu\nu}$.

As mentioned above, it has been argued that the Weyl-symmetric action, Eq. (2.1), 
is obtained in the case of a $\theta=0$ 
from the EH action by merely redefining the metric and scalar 
fields, that there are no conserved currents associated with the symmetry 
of Eq. (2.1), and that consequently this is a `sham' or `fake' Weyl-symmetry,  
e.g. [20-23]. 
Indeed, in the next section we illustrate this by going from the standard 
Friedmann-Robertson-Walker (FRW) 
action to Eq. (2.1) in the case $\theta=0$ by redefining fields. 
However, by adding another degree of freedom that is effectively only 
minimally-coupled to gravity (more specifically a free scalar field) to form 
a complex scalar field, 
the fields appearing in Eq. (2.1) cannot be generally redefined 
to recover the EH action; the two actions are clearly inequivalent in this case.
Here we only point out that integrating both Eqs. (2.2) \& (2.3) 
results in additional integration constants absent from GR. 
This will be addressed more concretely elsewhere [26].

\section{Cosmological Model}
In this section the background evolution, 
the evolution of linear perturbations, 
and a singularity-free early universe scenario are described.
The modulus of the complex scalar field, $\rho$, has essentially the same 
dynamics that the scale factor $a(\eta)$ has in the standard cosmological model 
(except near the turnaround point, 
where the dynamics of $\theta$ significantly alters that of $\rho$).
The field $\theta$ and its perturbations are responsible for the turnaround point 
and the flat power spectrum of density perturbations, respectively.
Under certain plausible conditions, the entire 
observable cosmic evolution, 
from BBN onward, is identical to that of the standard cosmological 
model, but very early universe processes and scenarios can be much 
different, e.g. there is no initial curvature singularity in 
the model, and an early inflationary period of evolution.

\subsection{Recasting FRW as a Scalar-Tensor Model and its Implications}
To motivate the construction in the following sections of a 
cosmological model based on Eq. (2.1) we work out 
the equivalence between the FRW action and Eq. (2.1)  
in the homogeneous and isotropic case in this section 
(in the case of identically vanishing $\theta$).

The issue of `wrong' relative sign of kinetic and potential 
energies as in Eq. (2.1) is characteristic of either general 
GR or other theories of 
gravitation `dressed' with conformal symmetry, e.g. [21, 23]. 
This is usually interpreted as an indication for instability of the 
theory, e.g. [28-30], and was at the heart of an intense 
debate regarding the physical equivalence of theories in 
the Einstein frame and Jordan frame where the latter was considered 
by many as `unphysical', in spite of the fact of being classically 
equivalent to the former, e.g. [31, 32], 
while others maintained that at least classically the 
theory with negative kinetic energy of the scalar field 
is stable. Indeed, the latter is the case with the classical FRW 
action, as we see below. Some even speculated that in a semi-classical context, 
when $\phi$ is quantized and the spacetime metric is treated as a classical 
field, stability may still be maintained, e.g. [33]. 
But, common lore holds it that if attempted to be quantized 
this `ghost' field causes a 
catastrophic production of particles due to the unlimited phase space 
available for such processes to take place. We thus conjecture that $\rho$ 
of Eq. (2.1) is unquantized, i.e. that it is always found in its particle 
vacuum state (its energy states are unquantized) -- 
a conjecture that could be better motivated by the following example. 
Again, we assume that $\rho$ of Eq. (2.1) is always classical and not 
only in the cosmological context [7].

Consider the EH action 
$\mathcal{I}_{EH}=(2\kappa)^{-1}\int (\tilde{R}-2\Lambda
+\tilde{\mathcal{L}}_{M})\sqrt{-\tilde{g}}d^{4}x$ 
with a cosmological constant $\Lambda$ and a matter Lagrangian 
density $\tilde{\mathcal{L}}_{M}$.
Here, $\kappa\equiv 8\pi G$ and $\tilde{R}$ is derived from the 
metric $\tilde{g}_{\mu\nu}$ in the usual way. 
The FRW action
\begin{eqnarray}
\mathcal{I}_{FRW}&=&(3/\kappa)\int[-a'^{2}+Ka^{2}
-\Lambda a^{4}/3+a^{4}\tilde{\mathcal{L}}_{M}(a)/6]\sqrt{-\tilde{g}}d^{4}x, 
\end{eqnarray}
is obtained for the metric 
$\tilde{g}_{\mu\nu}=a^{2}\cdot diag(-1,\frac{1}{1-Kr^{2}},r^{2},r^{2}\sin^{2}\theta)$, 
and after the term proportional to the corresponding curvature 
scalar $\tilde{R}=6(a''/a+K)/a^{2}$ is integrated by parts.
Here, $K$ is the spatial curvature, and 
a prime denotes derivatives with respect to conformal time, $d^{4}x=d^{3}xd\eta$.

One can readily verify that the Euler-Lagrange equation for the scale factor $a(\eta)$ 
that extremizes the action $\mathcal{I}_{FRW}$ is indeed the Friedmann equation 
$\mathcal{H}^{2}+K=a^{2}\tilde{\rho}_{M}+\frac{\Lambda a^{2}}{3}+const./a^{2}$, 
where $\mathcal{H}\equiv a'/a$ is the conformal Hubble function, 
$\tilde{\rho}_{M}\equiv -\frac{1}{6}\tilde{\mathcal{L}}_{M}$, and it is assumed that 
$\mathcal{\tilde{L}}_{M}(a)$ is a power-law in $a$, as $\tilde{\rho}_{M}$ 
is a power-law in standard cosmology. Completing the derivation 
requires that the integration constant is related to the energy density of 
radiation $\tilde{\rho}_{r}=const./a^{4}$. 

Defining $\rho^{2}\equiv 3a^{2}/\kappa$, the FRW action Eq. (3.1) 
is reformulated as a Weyl-symmetric scalar-tensor action with 
a non-positive kinetic term, defined on a static background 
\begin{eqnarray}
\mathcal{I}_{FRW}&=&\int\left(\frac{1}{6}R\rho^{2}-\rho'^{2}-V(\rho)
+\tilde{\mathcal{L}}_{M}(\rho)\right)\sqrt{-g}d^{4}x.
\end{eqnarray}
Here, $g_{\mu\nu}$ is the static metric related 
to the FRW metric via a Weyl transformation, 
$g_{\mu\nu}\equiv \tilde{g}_{\mu\nu}/a^{2}=diag(-1,\frac{1}{1-Kr^{2}},r^{2},r^{2}\sin^{2}\theta)$, 
the 4D infinitesimal volume element is, e.g. 
$\sqrt{-g}d^{4}x=\sinh^{2}\chi\sin\theta d\eta d\chi d\theta d\varphi$, 
where $\chi$ is a `radial' coordinate in the polar coordinate system 
(assuming $K<0$ for concreteness), $V(\rho)\equiv\lambda\rho^{4}$ is an effective potential, 
$\mathcal{L}_{M}=a^{4}\tilde{\mathcal{L}}_{M}$, $R=6K$ 
and $\lambda\equiv\kappa\Lambda/9$. 
The standard FRW spacetime is indeed described by a single degree 
of freedom, the scalar field $a(\eta)$, which is here replaced by $\rho(\eta)$, 
and therefore ignores the crucial role played by $\theta(\eta)$ as we will see below.

This example illustrates that a well-defined solution of the Einstein field equations 
is derived from an action which is essentially equivalent to Eq. (2.1), 
i.e. with a `wrong' relative sign of the kinetic and potential terms. 
This wrong sign appearing 
in Eq. (3.1) is never considered a problem ($a$ is part of the metric field) 
and similarly the sign of $\rho'^{2}$ should not alarm us, provided 
that $\rho$ is a genuinely classical field; same way that $a$ is a classical function 
in the standard formulation of the FRW solution so should be $\rho$.
We emphasize that unlike in the standard cosmological model where $\tilde{g}_{\mu\nu}$ 
and $\tilde{R}$ are evolving, the latter being divergent at the initial singularity, 
the curvature scalar $R=6K$ is static.
A Weyl transformation 
applied to Eq. (2.1) affects both the metric and scalar field to the extent 
that the dynamics of the metric (scale factor in our case) could be fully 
replaced by that of the scalar field in the 
FRW spacetime which is only described in terms of a 
single degree of freedom. 
In the standard cosmological model the kinetic and potential terms 
associated with the inflaton 
field appear with the canonical relative 
sign and thus could be quantized. 
In the present framework a different procedure is followed; 
whereas $\rho$, the analog of the scale factor $a$, is unquantized, 
the phase $\theta$ interacts only with the classical metric and $\rho$ fields 
and is therefore amenable to quantization.

Unlike $\rho(\eta)$ which is characterized by a vanishing current [23], 
the minimally-coupled field $\theta$ does have a conserved `charge' 
associated with it which plays a central role in the present 
cosmological model as is shown below. 
We express the view 
in this work that the scale factor $a$ in the GR-based formulation of the 
FRW cosmology is not more genuine or fundamental degree of freedom 
than the field $\rho$. 
Indeed, the scale factor $a(\eta)$ is not an observable and it is conventionally 
fixed to unity at present $a(\eta_{0})=1$, but its logarithmic derivative, 
the conformal Hubble function $\mathcal{H}=a'/a$, is. In analogy, 
$\mathcal{Q}\equiv\rho'/\rho$ is an observable in the present 
reformulation of the FRW model, and can be viewed as the rate of variation 
of particle masses and the Planck mass, i.e. $m'/m=O(10^{-18})\ {\rm sec^{-1}}$ 
at present, on a {\it static} background space as is described by Eq. (3.2). 

It is notable that the time coordinate in Eq. (3.2) 
is conformal time $\eta$ rather than cosmic time $t$. The two are 
related via $d\eta=dt/a(t)$ and as mentioned $a(t)$ is {\it arbitrarily} 
set to unity at present. In this particular (and standard) 
convention the cosmic and conformal `clocks' (that correspond to massive 
and massless test particles respectively) `tick' at the same rate at 
present. At any given time in the past ($a<1$) in the expanding space 
the conformal clock was ticking relatively slower, thereby explaining 
cosmological redshift. In contrast, in the present reformulation 
of the FRW action there is only a single clock -- the conformal 
clock -- and cosmological redshift is now explained by the dynamics 
of growing masses in the redshifting era (i.e. growing Rydberg `constant'). 

\subsection{Evolution of the Cosmological Background}

According to the alternative scenario proposed here a single 
complex scalar field 
does not only resolve the flatness and horizon puzzles, 
but also accounts for scale-free density perturbations. 
In this and the next section we derive the 
equations governing the background and linear perturbations, respectively.

The Einstein tensor components $G_{\mu}^{\nu}$, associated with 
the metric $g_{\mu\nu}=diag(-1,\frac{1}{1-Kr^{2}},r^{2},r^{2}\sin^{2}\theta)$,
in conformal rather than cosmic time coordinates, are $G_{\eta}^{\eta}=-3K$ and 
$G_{i}^{j}=-K\delta_{i}^{j}$.
In the following $f'\equiv\frac{\partial f}{\partial\eta}$ denotes the derivative of 
a function $f$ with respect to conformal time $\eta$.
Here, $i,j$ indices stand for the spatial coordinates. 
The energy-momentum tensor of a perfect fluid is 
$(T_{M})_{\mu}^{\nu}=\rho_{M}\cdot diag(-1,w,w,w)$, 
and employing Eq. (2.7) in Eqs. (2.1) \& (2.3) we obtain that 
\begin{eqnarray}
\mathcal{Q}^{2}+K&=&\frac{\rho_{M}}{\rho^{2}}-\theta'^{2}\\
2\mathcal{Q}'+\mathcal{Q}^{2}+K&=&-\frac{3w\rho_{M}}{\rho^{2}}+3\theta'^{2},
\end{eqnarray}
where $\mathcal{Q}\equiv\rho'/\rho$ is the conformal Hubble-like function.
In case that $\theta'=0$ these reduce to the FRW 
equations but with $a$ \& $\mathcal{H}$ replaced by $\rho$ \& $\mathcal{Q}$. 
This implies that in the radiation-dominated (RD) or 
matter-dominated (MD) evolutionary eras, 
$\rho\propto\eta^{2/(1+3w)}$ is a monotonically 
increasing function of conformal time. This continues 
to be the case insofar $w>-1/3$. 
The evolution of the Rydberg `constant' thus explains the observed 
cosmological redshift on this {\it static} spacetime. 
Since $\rho(\eta)$ is governed by a Friedmann-like equation it then follows that 
the effective background metric `seen' by a massive test particle of mass $m$ is 
$m^{2}g_{\mu\nu}\propto\rho^{2}\eta_{\mu\nu}\propto a^{2}\eta_{\mu\nu}$ which 
is exactly the FRW metric. Massless test particles 
follow geodesics of a Minkowski 
background metric, as in the standard cosmological model (with $K=0$).

Before exploring the profound 
ramifications on the background evolution entailed by introducing the field $\theta$, 
we comment on the background 
evolution in the observable redshifting universe.
We emphasize that exactly as in the standard 
cosmological model, $\rho_{M}$ is the sum of various contributions, 
i.e. $\rho_{M}=\sum_{i}\rho_{i}$ where 
each $\rho_{i}$ scales according to Eq. (2.7) with its respective $w_{i}$. 
In the same fashion that in the 
standard cosmological model the energy density associated with a species of fixed EOS 
$w_{i}$ is $\rho_{i}\propto a^{-3(1+w_{i})}$, i.e. the ratio of energy densities 
associated with two species which are characterized by $w_{i}$ \& $w_{j}$ is $a^{-3(w_{i}-w_{j})}$, 
so it is the case here where $\rho_{i}\propto \rho^{1-3w_{i}}$ and the ratio becomes 
$\rho^{-3(w_{i}-w_{j})}$, consistent with the fact that $\rho(\eta)$ in the present model 
replaces $a(\eta)$ of the standard cosmological model. 
The sequence of RD, MD redshifting 
followed by the recent vacuum-like dominated 
redshifting is exactly as in the standard cosmological model. 
Clearly, since 
space is static (and assuming adiabatic evolution where 
particles are neither annihilated nor produced) particle number densities 
are fixed and the entire evolution is governed by the evolution of particle masses, which 
in turn is regulated by $\rho$. In other words, the monotonic growth of $\rho$, i.e. of particle 
masses, in a static background space (where the energy density 
of the CMB, i.e. its temperature, is fixed) implies 
that, e.g. an emitted atomic line at earlier epochs is determined by a relatively 
larger Rydberg `constant' and is correspondingly characterized by a longer wavelength, i.e. 
the observed cosmological redshift simply reflects the monotonic growth of particle masses 
over cosmological time scales. 

In addition, it is easy to see that the microphysics of 
the early universe is identical to that of the standard cosmological model. 
For example, the Saha equation
\begin{eqnarray}
\frac{x_{e}^{2}}{1-x_{e}}=\frac{45s}{4\pi^{2}}\left(\frac{m_{e}}{2\pi T}\right)^{\frac{3}{2}}
\exp(-\Delta/T), 
\end{eqnarray}
determines the temperature $T$ at recombination, where $x_{e}$ is the ionization 
fraction, $s=(\frac{4\pi^{2}}{45})\frac{T^{3}}{n_{b}}$ 
and $n_{b}$ is the baryon number density, and $\Delta\equiv\alpha_{e}^{2}m_{e}/2$ 
with $\alpha_{e}$ \& $m_{e}$ being the fine structure constant and 
the electron mass, respectively. 
In the standard cosmological model $m_{e}=constant$ \& $T\propto a^{-1}$ and consequently 
the terms $\Delta/T$, $\left(\frac{m_{e}}{2\pi T}\right)^{\frac{3}{2}}$ \& $s$ 
scale $\propto a$, $a^{3/2}$ \& $a^{0}$, respectively. For comparison, in the present model 
$m_{e}\propto\rho(\eta)$ \& $T\propto \rho^{0}$, and thus $\Delta/T$, 
$\left(\frac{m_{e}}{2\pi T}\right)^{\frac{3}{2}}$ \& $s$ 
scale $\propto\rho$, $\rho^{3/2}$ \& $\rho^{0}$, respectively. Consequently, the physics 
of recombination in the proposed model 
is identical to that of the standard cosmological model. Similar arguments 
apply to the physics of BBN which remains unchanged with respect to the standard 
cosmological model. 

Another simple example is the Compton scattering cross-section in the Thomson limit, 
$\tau=\int n_{e}\sigma_{T}dl$, where $\sigma_{T}\propto m_{e}^{-2}$ is the cross section 
for Compton scattering in this limit. 
In the standard cosmological model $n_{e}\propto a^{-3}$, $\sigma_{T}\propto a^{0}$ 
\& $dl\propto dt\propto a(\eta) d\eta$, i.e. $\tau=\int a^{-2}(\eta)d\eta$. In comparison, 
in the proposed model $n_{e}\propto a^{0}$, $\sigma_{T}\propto a^{-2}$ 
\& $dl\propto d\eta$, i.e. $\tau=\int \rho^{-2}(\eta)d\eta$.

Back to the `Friedmann equations', Eqs. (3.3) \& (3.4). 
These are augmented with $\propto\theta'^{2}$ terms,    
which come from $\Theta_{\mu}^{\nu}$, the effective energy-momentum tensor 
associated with the kinetic term of the scalar field (Eq. 2.5). 
These terms are comparable in size (but are of opposite sign) 
to other contributions to the energy budget of the 
universe only close to the turnaround time 
and are negligible at all 
other times as is shown below.
Whereas $\rho(\eta)$ essentially replaces the 
scale factor $a(\eta)$ of the FRW model, $\theta$ is a new field. 
As is already mentioned and as shown in 
this section and in 3.4, inclusion of a non-vanishing $\theta'$ in our cosmological model 
serves two purposes. First, it could be used to avoid the initial singularity 
by explicitly violating the weak, strong, and null energy conditions. 
Second, it couples excitations of $\theta$ to scalar metric perturbations, i.e. 
to the gravitational potential, 
ultimately resulting in a flat power spectrum. This mechanism has no parallel in the 
tensor perturbations sector and consequently no primordial 
generation of B-mode polarization on cosmological scales is 
expected (see also section 3.4).  

Now, to close our system of background equations, 
the evolution of $\theta$ is governed by
\begin{eqnarray}
\theta''+2\mathcal{Q}\theta'=0,
\end{eqnarray}
i.e. $\theta'=c_{\theta}/\rho^{2}$, where $c_{\theta}$ is an integration constant.
Eq. (3.6) is obtained for the second 
component of $\phi=(\rho,\theta)$ from Eq. (2.3) 
with $\mathcal{G}_{IJ}=2\cdot diag(-1,-\rho^{2})$, 
or is alternatively derived directly from variation 
of the kinetic term in Eq. (2.1) with respect to $\theta$ 
because all other terms are independent of $\theta$ and its derivatives. 
Another way to obtain Eq. (3.6) is to employ 
Eq. (2.3) to the complex field frame described below Eq. (2.1) while taking 
advantage of the fact that both $F$, $V$ \& $\mathcal{L}_{M}$ 
depend purely on $|\phi|$ and that $\mathcal{G}_{IJ}=constant$ in this frame. 
The resulting equation is $\phi^{*}\frac{\partial\mathcal{I}}{\partial\phi^{*}}
-\phi\frac{\partial\mathcal{I}}{\partial\phi}=2{\rm Im}(\phi^{*}\Box\phi)=0$ 
from which Eq. (3.6) readily follows.

Similar to the discussion below Eq. (3.1), the field 
equation Eqs. (3.3), (3.4) \& (3.6) could be directly derived from variation of 
$\mathcal{I}=\int\left(K\rho^{2}-\rho'^{2}-\rho^{2}\theta'^{2}
-\rho_{M}(\rho)\right)\sqrt{-g}d^{4}x$ with respect to $\rho$ \& $\theta$ 
while making use of Eq. (2.7). The metric $g_{\mu\nu}$ is the same static 
metric field used in Eq. (3.2).
Substituting $\theta'=c_{\theta}/\rho^{2}$ into Eqs. (3.3) \& (3.4), and 
comparing with Eq. (2.7), shows that the $\propto\theta'^{2}$ terms 
effectively play the role of a `stiff matter' contribution 
with $w_{\theta}=1$. This contribution to the energy budget 
would have dominated the cosmic evolution 
at early epochs, i.e. in the small field limit $\rho/\rho_{0}\ll 1$, 
has it been positive. However, from Eq. (3.3) it is clear that 
$\rho_{\theta}\equiv -\rho^{2}\theta'^{2}$ is negative. 
Assuming no turnaround takes place during the MD era, and 
if it ever takes 
place it does so deep into the RD era, then only when $|\rho_{\theta}|$ 
becomes comparable to the energy density of radiation does 
a turnaround take place, after 
which $|\rho_{\theta}|$ drops faster than any other form of energy 
density considered in this model in the redshifting era  
(we assume a theoretical upper limit of $w\leq 1/3$ for perfect fluids). 
Eq. (2.3) also gives a second-order evolution equation 
for $\rho$, but employing $R=6K$ and the solution of Eq. (3.6), 
$\theta'=c_{\theta}/\rho^{2}$, this equation can be readily integrated once. 
The result of this integration is exactly the Friedmann-like equation, 
Eq. (3.3), and therefore the system of 
Eqs. (3.3), (3.4) \& (3.6) fully determine the background 
evolution on cosmological scales. It should 
be mentioned that the relation $\theta'=c_{\theta}/\rho^{2}$ 
has been similarly obtained in [34] in the context of another 
$U(1)$ symmetric cosmological model which is otherwise 
entirely different from the model put forward in the present work.

The proposed scenario is non-singular and the cosmic history comprises 
of blue- and redshifting evolution epochs. 
The former, blueshifting epoch, is nearly symmetric in this scenario 
to the redshifting era. 
It starts with a vacuum-like energy-, 
followed by MD, RD, and a brief era dominated by a mixture of radiation 
and `stiff' energy density, a turnaround, 
and a redshifting epoch with these 
various eras occurring in reverse order. 

Accounting for the vacuum-like, NR, radiation and effectively stiff 
energy densities, Eq. (3.3) implies that 
\begin{eqnarray}
\mathcal{Q}^{2}=\lambda\rho^{2}+\lambda_{p}n_{NR}/\rho+\rho_{r,*}/\rho^2+\rho_{\theta,*}/\rho^{4},
\end{eqnarray}
where $\rho_{i,*}$ is the energy density associated with the i'th species 
at $\eta_{*}$, $\rho(\eta_{*})\equiv 1$, $n_{NR}$ is the number density of NR particles (assuming 
a single species for illustration purposes), and $\lambda_{p}$ is a dimensionless parameter.
In a static background, as the one considered here, both $n_{NR}$ and $\rho_{r}$ 
are fixed constants.
The full analytic integration of this equation is not very illuminating, and therefore we 
treat two interesting limits separately that will suffice for our purposes.  
The first limit is obtained by neglecting the vacuum-like and NR terms 
near the turnaround point. 
In this case, the Friedmann-like equation integrates to
\begin{eqnarray}
\rho^{2}=\rho_{r,*}\eta^{2}-\rho_{\theta,*}/\rho_{r,*}
\end{eqnarray}
where $\rho$ attains its minimum at $\eta=0$.
It could be readily integrated to give the cosmic time around the turnaround point, 
$t\propto\int \rho(\eta)d\eta$, and is easily verified to be non-singular as well. 
In other words, the effective time coordinates of both massless and massive particles 
can be extended through the turnaround point, 
i.e. spacetime is geodesically-complete as is generically the 
case in non-singular bouncing cosmological models, [35]. 
In the absence of turnaround point 
(e.g. in case $\theta$ identically vanishes) the scalar field would have scaled 
$\rho\propto\eta$ as does the scale factor $a(\eta)$ in the RD era. 
The scalar field 
then vanishes at $\eta=0$. Whereas this is not a curvature singularity, it is 
a topological singularity and it is not entirely clear to us that the theory 
can be extended to $\eta<0$ as this would (at least naively) 
imply negative masses. 
Incoming null geodesics are described by $\eta_{0}-\eta=r$ in 
Minkowski spacetime. 
Therefore, since $\eta$ is bounded from below this would imply 
that observable $r$ is finite although underlying 
spacetime is Minkowski; this happens in that case because 
scalar fields hit an `initial (topological) singularity' 
(in the scalar field $\rho$, not the metric or curvature scalar which are trivial) 
at a certain finite time in the past.

In the other extreme -- where the background dynamics is dominated by the 
quartic potential -- $\rho=\left(\sqrt{\lambda}(\eta_{c}-\eta)\right)^{-1}$, 
and $\rho$ scales according to its canonical dimension $length^{-1}$, 
i.e. $\propto\eta^{-1}$, not $\propto t^{-1}$. 
This again highlights the privileged 
role played by conformal time as compared to cosmic time, in contrast to the standard 
cosmological model where conformal time is only used for computational convenience, 
or as the natural time coordinate parameterizing null geodesics. 
The integration constant $\eta_{c}$ determines the lower and upper limit on 
the (conformal) time coordinate in the proposed nearly symmetric model, 
$\eta\in(-\eta_{c},\eta_{c})$. Again, since $\eta$ is bounded 
from below, this time due to the presence of the 
vacuum-like energy density component, 
then observed radial distances are bounded 
at $\eta=-\eta_{c}$ where the scalar field 
diverges and the model breaks down. Specifically, null geodesics in the case $K=0$ 
satisfy $r(\eta)=\eta_{0}-\eta$. Since $\eta>-\eta_{c}$ then the maximal 
observable distance is $r_{max}=\eta_{0}+\eta_{c}$. The latter can potentially 
be much larger than the Hubble scale if $\eta_{c}\gg\eta_{0}$. 
Although we focus here on DE-dominated asymptotics 
it is clear that a similar breakdown of the model is shared by 
any model asymptotically dominated by $w<-1/3$ as Eq. (3.3) 
integrates in this case to $\rho=c\left(\eta_{c}\pm\eta\right)^{\frac{2}{1+3w}}$ 
(in the cases $\eta<0$ \& $\eta>0$ respectively) where $c$ and $\eta_{c}$ are 
positive integration constants. Note that this is also the solution of the 
(equivalent) Friedmann equation, the scale factor $a(\eta)$. Consequently, 
this scalar field singularity is present in the standard cosmological model 
as well. However, in the latter it is only a future singularity, whereas 
in the proposed model it is also a past singularity taking place during 
the blueshifting epoch, thereby rendering the {\it observable} 
universe finite although Minkowski space itself has no singularity 
and therefore no horizon associated with it. 

One may argue that instead of applying Eq. (2.1) to a homogeneous 
and isotropic spacetime as done in the present section, we could have equally 
well used the EH action, Eq. (3.1), supplemented by 
$\mathcal{I}_{\theta}=-(3/\kappa)\int\theta^{\mu}\theta_{\mu}\sqrt{-g}d^{4}x
=(3/\kappa)\int a^{2}\theta'^{2}d^{4}x$, in formulating 
the cosmological model described in the present section on an expanding 
background space. However, this {\it ad hoc} procedure -- while being 
mathematically equivalent to the approach taken here -- seems somewhat 
less natural than the $U(1)$ symmetric model outlined here and 
formulated on a static background space.

The `flatness problem' arises in the hot big bang model due to the monotonic 
expansion of space and the consequent faster dilution of the energy density 
of matter (either relativistic or NR) compared to the effective energy 
density dilution associated with curvature. It is thus hard 
to envisage how could space be nearly flat (as is indeed inferred from 
observations, e.g. [36]) 
if not for an enormous fine-tuning at the very early universe, or an 
early violent inflationary era. 
In the proposed scenario the matter content of the universe 
has always been the same, 
and in particular the present ratio of matter- to curvature-energy densities 
has been exactly the same when $\rho$ in the 
blueshifting era was equal to its 
present (redshifting era) value. 
However, in the blueshifting era matter domination over curvature 
is actually an {\it attractor point} as the blueshifting universe starts essentially 
from $\rho\rightarrow\infty$. 
In other words, had the universe been curvature-dominated (CD) 
at present (as is naively expected in the standard 
{\it expanding} hot big bang model but with no inflation), 
i.e. at $\rho_{0}$, it must have been CD 
at $\rho_{0}$ at the mirror blueshifting era, 
but since $\rho_{M}\propto\rho$ while $\rho_{K}\propto\rho^{2}$ 
then a curvature domination at $\rho_{0}$ in blueshifting era 
would amount to an extremely fine-tuned 
$\rho_{M}/\rho_{K}\rightarrow 0$ at $\rho\rightarrow\infty$ 
at the blueshifting era.
As is well-known, entropy produced in the pre-bounce era could be processed 
at the bounce to thermal radiation, implying in effect 
that $\rho_{r}$ might somewhat 
change between pre- and post-bounce but the expectation in the proposed model 
is that the contracting and expanding epochs nearly mirror each other. 

The kind of argument employed here in explaining away the flatness problem 
can be reversed to show that anisotropy 
actually does enormously grow relative to the other energy 
species in the blueshifting era. 
In terms of standard cosmology $\rho$ represents some (geometric) average scale factor 
and $\theta$ determines deviation from isotropy, such as in Bianchi-type cosmological models. 
In bouncing models this anisotropy is a measure of the variation 
of expansion or contraction rates between the principal 
axes of the homogeneous model, 
which change very fast near the bounce, the Belinskii-Khalatnikov-Lifshitz (BKL) 
instability, i.e. that anisotropy 
is a natural attractor in contracting cosmologies, and in order to avoid it an enormous 
fine-tuning of initial conditions has to be invoked, e.g. [37]. 
The analog of this in the proposed model would be the 
extreme variation over time of $\theta$ around the turnaround point. 
We will argue below that our model is free of these illnesses that generically 
afflict bouncing cosmological models. It should be emphasized that slightly 
anisotropic expansion or contraction could be mimicked by a stiff 
matter but the opposite 
is not true -- an effective stiff EOS does not necessarily imply 
anisotropic evolution. For example, in the proposed model space 
is static and isotropic 
and the evolution is only in the scalar field, not space. 
An effective anisotropy, or in other words, an effective `stiff' component 
($w=1$) could arise, e.g. from terms in the Lagrangian Eq. (2.1) which 
are $\propto|\phi|^{-2}$. However, Weyl symmetry that prohibits the appearance 
of any dimensional quantity at the action level, severely 
limits the existence of such a term. It could appear in the form, 
e.g. $\Delta\mathcal{L}=(\bar{\psi}\psi)^{2}|\phi|^{-2}$, 
or higher-curvature terms of the form 
$\propto|\phi|^{-2}C_{\alpha\beta}^{\gamma\delta}C_{\gamma\delta}^{\rho\sigma}
C_{\rho\sigma}^{\alpha\beta}$. The second term vanishes in the 
case of conformally-flat metric $g_{\mu\nu}$, as the one employed here. 
Perhaps more important, in the same fashion that 
no stiff matter component seems to be required in standard cosmology to explain 
observations we ignore such terms in the proposed model, and in general 
do not allow non-canonical negative powers of the scalar field to appear 
in the action (as discussed below Eq. 2.7) for a lack of an otherwise good 
reason to allow such terms.

\subsection{Linear Perturbation Theory}
The standard cosmological model has successfully passed numerous tests and has been quite 
effective in explaining the formation and linear growth of density perturbations over the 
background spacetime, predicting the CMB acoustic peaks, polarization spectrum, and 
damping features on small scales. It also correctly describes the linear and nonlinear 
evolution phases of the LSS (on sufficiently large scales) and abundance of 
galaxy and galaxy cluster halos. 
Therefore, it would seem essential to establish equivalence of linear 
perturbation theory between our model and the standard cosmological model. 

As in section 3.2, we assume an effective energy density $\rho_{M}$ 
characterized by a (generally time-dependent) EOS $w=w(\eta)$ 
that encapsulates NR and relativistic baryons, CDM, 
radiation, and a vacuum-like energy density. 
Two perturbation variables are the scalar metric perturbations 
$\varphi$ \& $\alpha$ that appear in the rescaled perturbed FRW line element
$ds^{2}=-(1+2\alpha)d\eta^{2}+(1+2\varphi)\gamma_{ij}dx^{i}dx^{j}$, 
where $\gamma_{ij}\equiv diag[1/(1-Kr^{2}),r^{2},r^{2}\sin^{2}\theta]$.
We define the fractional energy density and pressure perturbations 
(in energy density units) $\delta_{\rho_{M}}\equiv\delta\rho_{M}/\rho_{M}$ and 
$\delta_{P_{M}}\equiv\delta P_{M}/\rho_{M}$, respectively. 
The matter velocity is $v$. 

In the shear-free gauge, and in the case of vanishing 
curvature $K$ and stress anisotropy, i.e. $\varphi=-\alpha$, 
the Arnowitt-Deser-Misner (ADM) energy \& momentum constraints, Raychaudhuri equation, 
$\delta\theta$ equation, and the perturbed continuity \& Euler equations 
(Eqs. 39, 40, 42, 43, 48 \& 49 of [24]) reduce to
\begin{eqnarray}
&&\mathcal{Q}\tilde{\varphi}'+\left(\frac{k^{2}}{3}+\mathcal{Q}^{2}
+\theta'^{2}\right)\tilde{\varphi}=-\frac{1}{2}(\mathcal{Q}^{2}+\theta'^{2})\tilde{\delta}_{\rho_{M}}-\theta'\delta\theta'\\
&&\tilde{\varphi}'+\mathcal{Q}\tilde{\varphi}=-\frac{3(1+w)}{2}(\mathcal{Q}^{2}+\theta'^{2})u+3\theta'\delta\theta\\
&&\tilde{\varphi}''+2\mathcal{Q}\tilde{\varphi}'+\left[2\mathcal{Q}'-4\theta'^{2}
-\frac{k^{2}}{3}\right]\tilde{\varphi}=\nonumber\\
&&\frac{(1+3w)}{2}(\mathcal{Q}^{2}+\theta'^{2})\tilde{\delta}_{\rho_{M}}+4\theta'\delta\theta'\\
&&\delta\theta''+2\mathcal{Q}\delta\theta'+k^{2}\delta\theta=-4\theta'\varphi'\\
&&\tilde{\delta}'_{\rho_{M}}+(1+w)(3\tilde{\varphi}'+k^{2}u)=0\\
&&u'+(1-3w)\mathcal{Q}u+\frac{w'u}{1+w}=-\tilde{\varphi}+\frac{w\tilde{\delta}_{\rho_{M}}}{1+w}\quad ,
\end{eqnarray}
where $u\equiv v/k$ and we used Eq. (3.3) to eliminate $\rho_{M}$, 
and the `shifted' perturbation quantities are defined as  
$\tilde{\varphi}\equiv \varphi+\delta_{\rho}$ \& 
$\tilde{\delta}_{\rho_{M}}\equiv\delta_{\rho_{M}}-4\delta_{\rho}$, 
where $\delta_{\rho}\equiv\frac{\delta\rho}{\rho}$. Applying the conformal transformation 
$g_{\mu\nu}\rightarrow g_{\mu\nu}/\rho^{2}$ \& $\rho_{M}\rightarrow\rho_{M}\rho^{4}$ 
takes us back to $\varphi$ and $\delta_{\rho_{M}}$ in the limit of linear 
perturbations $\delta_{\rho}\ll 1$.
We note that the difference of the second-order perturbation 
equation for $\delta\rho$ and the perturbed trace of the 
generalized Einstein equation (Eqs. (43) \& (44) of [24], respectively) 
results in the consistency relation $\delta_{P_{M}}=w\delta_{\rho_{M}}$ and 
therefore provides no additional information to Eqs. (3.9)-(3.14).
The latter are exactly the 
linear perturbation equations over an FRW background if 
we make the replacement $\mathcal{Q}\rightarrow\mathcal{H}$ and recalling 
the conclusion from section 3.2 that $\mathcal{Q}$ is exactly $\mathcal{H}$ 
except of very near the turnaround point. 

Vector and tensor perturbations, which are described by Eqs. (53)-(55) 
and (58) of [24] respectively, similarly satisfy the same equations 
they do in GR, provided that $a\rightarrow\rho$, 
i.e. $\mathcal{H}\rightarrow\mathcal{Q}$ in going from the GR 
formulation to the one adopted here, and the sources, e.g. 
anisotropic stress $\pi^{(t)}$, are correspondingly rescaled by 
a multiplicative factor $\rho^{4}$, 
e.g. $\pi^{(t)}\rightarrow\pi^{(t)}\rho^{4}$.
For example, the vector and tensor perturbation modes, 
$v^{(v)}$ \& $h$ respectively, satisfy the equations
\begin{eqnarray}
[a^{4}\rho_{M}(1+w)v^{(v)}]'&=&-\frac{(k^{2}-2K)}{2k}a^{4}\pi^{(v)}\nonumber\\
h''+2\mathcal{H}h'+(k^{2}+2K)h&=&8\pi Ga^{2}\pi^{(t)},
\end{eqnarray}
in the standard cosmological model, where $\pi^{(v)}$ \& $\pi^{(t)}$ are the vector 
and tensor anisotropic stresses, respectively. For comparison, the 
corresponding Eqs. (54) \& (58) of [24] are
\begin{eqnarray}
[\rho_{M}(1+w)v^{(v)}]'&=&-\frac{(k^{2}-2K)}{2k}\pi^{(v)}\nonumber\\
h''+2\mathcal{Q}h'+(k^{2}+2K)h&=&\frac{3\pi^{(t)}}{\rho^{2}},
\end{eqnarray}
respectively. For example, the decay of superhorizon tensor modes, $h\propto a^{-2}$, 
in the standard cosmological model is replaced by the identical 
scaling $h\propto\rho^{-2}$ in the alternative model discussed here. This is not surprising; 
the two cosmological models are related via Weyl symmetry (ignoring the effect of the dynamical field 
$\theta$ near the turnaround point for 
the purpose of the present discussion) 
and thus the dynamics of dimensionless metric perturbations 
such as $v^{(v)}$ \& $h$ should be identical in both models.

The full kinetic theory, pertaining to the theory described by Eq. (2.1) 
applied to homogeneous and isotropic background cosmology, 
involving collisional photons and collisionless 
neutrinos, e.g. [24], where the corresponding perturbed energy-momentum tensor 
components are given in terms of integrals over the respective 
distribution functions $f$, can be easily incorporated in our scheme 
with minor adjustments; neutrino masses are $\propto\rho(\eta)$, the dynamical 
Higgs VEV, and this has to be accounted for in the collisional 
Boltzmann equation.

The Newtonian limit of the gravitational interaction in the framework of Eq. (2.1)
is obtained from Eqs. (3.9)-(3.14) by setting $\mathcal{Q}$, $K$, $w$, 
and $\theta'$ to zero. In particular, Eqs. (3.9), (3.13) \& (3.14) are the relativistic 
Poisson, continuity, and Euler equations, respectively. 

In the standard cosmological model, 
a gravitational potential consistent with Eqs. (3.9)-(3.14) 
in the RD era, i.e. $w=1/3$, 
is $\varphi=A(k)j_{1}(q)/q+B(k)y_{1}(q)/q$ where $q\equiv \frac{x}{\sqrt{3}}$, 
with $x\equiv k\eta$, and $j_{l}$ \& $y_{l}$ are the spherical 
Bessel and Neumann functions, respectively.
Assuming the two modes have been generated at approximately equal amplitudes at 
some $q\ll 1$ during the inflationary period 
then the mode that 
diverges at $q=0$ is negligible at later times. 
Consequently, the initial condition $B(k)=0$ is selected 
in standard cosmology and $\varphi$ is essentially 
constant at $q\rightarrow 0$. In the present model the argument is different; 
since the model is extended through the turnaround point to $\eta<0$ the 
diverging mode 
which is an odd function of $\eta$ near the turnaround point is set to zero by the mere 
requirement that $\varphi$ is a continuous function of time, at $\eta=0$ in particular. 
Hence, the adiabatic initial condition merely follows from the very 
existence of a non-singular turnaround and required continuity of the 
gravitational potential. 
A similar argument applies to the tensor modes; the mode singular at $\eta=0$ 
is an odd function of $\eta$ and is consequently discontinuous at $\eta=0$.
Although the simplest models of 
inflation guarantee that the initial conditions are adiabatic 
in the standard cosmological model, a mixture of adiabatic and isocurvature 
modes can also be accommodated by certain inflationary models. 
The latter are constrained at the 
few percent level but are not entirely ruled out [36].

It has been recently proposed that the universe has 
a CPT `anti-universe' counterpart [38].
The two universes according to this picture share a common topological 
singularity at $\eta=0$. 
In contrast, as is shown in section 3.4 below, our model is 
characterized by a non-singular turnaround. In [38] 
it is shown that the adiabatic initial conditions could naturally 
result from time-reversal symmetry which is inherent to their model, 
a conjecture that is obviously not satisfied by the perturbed 
FRW universe in the scenario proposed here (but is clearly satisfied 
at the background level of the present model).
Thus, rather than conjecturing a {\it global} time-reversal symmetry 
we make a more modest and natural requirement from our model -- 
pointwise continuity. 
This implies in particular that the integration 
constants multiplying perturbation modes which are 
divergent odd functions of $\eta$ must vanish. 

\subsection{Primordial Flat Spectrum and Turnover Point}

Although inflation 
provides a mechanism for generating scalar and tensor 
perturbations which are characterized 
by nearly-flat power spectra, it is not a {\it prediction} of 
the inflationary scenario; it 
has been known for nearly a decade before the advent of inflation 
that at least the density 
perturbations are described by a nearly flat spectrum [39-41], and 
as of yet there is no evidence for the existence of primordial tensor 
modes anyway. 
Other early universe scenarios, e.g. the varying speed of light 
cosmology [42], the ekpyrotic [43] and new ekpyrotic [44] 
scenarios, the cyclic universe [45, 46], string gas cosmology [47], 
Anamorphic cosmology [48], and pseudo-conformal universe [49, 50], are 
capable of explaining the observed flat spectrum as well.

It is well known that flat spectra are equally well 
generated from fluctuations of scalar fields during 
the MD contraction era in bouncing scenarios [51]. 
Alternatively, here we consider the massless transversal 
perturbation of the scalar field, $i\rho\delta\theta$  
(accounting for its minimal 
coupling to scalar curvature perturbations) 
as a viable source for scale-invariant density perturbations. 

Combining Eqs. (3.9) \& (3.11) we obtain for an arbitrary $w(\eta)$
\begin{eqnarray}
\varphi''+3(1+w)\mathcal{Q}\varphi'+wk^{2}\varphi=3(1-w)\theta'
\delta\theta'.
\end{eqnarray}
Using Eq. (3.12) to express $\varphi'$ in terms of $\delta\theta$ 
and its derivatives, taking the time-derivative of Eq. (3.17), 
and employing the relation $\theta'\propto\rho^{-2}$, 
i.e. $\theta''/\theta'=-2\mathcal{Q}$, and $\mathcal{Q}=\frac{2}{(1+3w)\eta}$ 
(by virtue of Eq. 3.3 and assuming 
the dynamics is dominated by a species 
characterized by $w=constant$, a very good approximation throughout the cosmic 
history except at very brief transitions between the various epochs), 
we obtain a fourth-order equation for $\delta\theta$ of the form 
$\delta\theta_{xxxx}+f_{3}\delta\theta_{xxx}+f_{2}\delta\theta_{xx}+f_{1}\delta\theta_{x}
+f_{0}\delta\theta=0$ with
\begin{eqnarray}
f_{0}&\equiv&\left[\frac{32}{(1+3w)^{2}}-2\right]\frac{1}{x^{2}}+w\nonumber\\
f_{1}&\equiv&\frac{24(3+w)(1-3w)}{(1+3w)^{3}x^{3}}+\frac{2(7+5w)}{(1+3w)x}\nonumber\\
f_{2}&\equiv& 1+w+\left[\frac{80}{(1+3w)^{2}}-2\right]\frac{1}{x^{2}}\nonumber\\
f_{3}&\equiv&\frac{6(3+w)}{(1+3w)x},
\end{eqnarray}
where again $x\equiv k\eta$, and small $\propto\theta_{x}^{2}$ terms 
have been neglected. The latter is a very good approximation except 
for the immediate vicinity of the turnaround point.
We have not been able to analytically solve Eq. (3.18) 
for arbitrary $w$. Specializing to the case $w=0$ results in
\begin{eqnarray}
\delta\theta&=&\frac{1}{x^{3}}\left[c_{1}+c_{2}e^{ix}(1-ix)+c_{3}e^{-ix}(1+ix)\right]\nonumber\\
&+&\frac{c_{4}e^{-ix}}{x^{8}}\left[e^{2ix}(1-ix)x^{5}{\rm Ei}(-ix)\right.\nonumber\\
&+&\left.i\left(4e^{ix}(72-4x^{2}+x^{4})+x^{5}(x-i){\rm Ei}(ix)\right)\right],
\end{eqnarray}
where $c_{1}$, $c_{2}$, $c_{3}$ \& $c_{4}$ are 
(possibly $k$-dependent) integration constants, and Ei is the Exponential Integral.
As usual, we impose the Bunch-Davies condition in the limit $x\gg 1$ 
on the perturbed transversal component of the doublet, 
i.e. $i\rho\delta\theta\rightarrow\frac{1}{\sqrt{2k}}e^{-ix}$. 
Employing $\rho=\rho_{0}(\eta/\eta_{0})^{2}$ in the MD era then 
implies that $c_{3}=-\frac{k^{3/2}\eta_{0}^{2}}{\sqrt{2}\rho_{0}}$ and all other 
integration constants vanish. Using 
Eq. (3.6), $\theta'=c_{\theta}/\rho^{2}$, in Eq. (3.17) we obtain in the limit $x\ll 1$
\begin{eqnarray}
\varphi''+\frac{6\varphi'}{\eta}=\frac{C}{\eta^{8}}, 
\end{eqnarray}
where $C\equiv(-9c_{\theta}\eta_{0}^{6})/(\sqrt{2}k^{3/2}\rho_{0}^{3})$.
Its solution is $\varphi=\tilde{c}_{1}+\tilde{c}_{2}\eta^{-5}+(C/6)\eta^{-6}$. 
In the blueshifting era the fastest growing mode is the $\propto \eta^{-6}$ 
term (assuming all modes are generated at approximately similar amplitudes, 
and neglecting the $\propto\eta^{-5}$ term that is discontinuous at $\eta=0$) 
and thus $\varphi=-9c_{\theta}/(6k\sqrt{2k}\rho^{3})$, 
which implies that
\begin{eqnarray}
\Delta^{2}=\frac{k^{3}|\varphi^{2}|}{2\pi^{2}}=\frac{9\lambda|\rho_{\theta}|}{16\pi^{2}\rho_{DE}} 
\end{eqnarray}
is flat. Perturbations generated earlier on during the MD blueshifting era 
are relatively smaller at the time of production than those which are generated latter 
since $\rho_{\theta}$ is much smaller than $\rho_{DE}$ at large $\rho$ values. 
However, they have 
a longer time to grow than the perturbations generated 
closer to the turnaround point. 
Overall, the perturbations observed today are equally contributed 
at all times during blueshifting MD era.

It should be stressed that unlike the mechanism proposed in [51] 
which reflects the fact that any field fluctuates on an evolving background 
(and with rms fluctuation determined by the Hubble scale), 
the curvature perturbations discussed here are a direct result of the coupling between $\theta$ 
and $\varphi$ (and the amplitude of scalar perturbations is therefore $\propto\rho_{\theta}$). 
There is no tensorial analog in our scenario to this coupling.

Repeating the same procedure in the RD era ($w=1/3$), and imposing the 
Bunch-Davies vacuum on $i\rho\delta\theta$ in the $x\gg 1$ limit 
eliminates one of the four integration constants. In the other extreme, 
$x\ll 1$, we obtain $\delta\theta\propto\eta^{-4}$. 
Since $\theta'=c_{\theta}{\rho^{2}}$ 
and since in the RD era $\rho=\rho_{0}\eta/\eta_{0}$, then neglecting 
the $wk^{2}\varphi$ term in Eq. (3.17) we obtain 
\begin{eqnarray}
\varphi=\frac{\beta}{\eta^{5}}+\frac{c'_{1}}{\eta^{3}}+c'_{2} 
\end{eqnarray}
in the $x\ll 1$ limit, where $c'_{1}$, $c'_{2}$ \& $\beta$ 
could possibly depend on $k$. 
The constant $\beta$ 
is also proportional to $c_{\theta}$. Again, the mere requirement 
of continuity at all times, and at $\eta=0$ in particular, implies that 
$\varphi=c'_{2}(k)$, and consequently Eq. (3.21) is not modulated during 
the RD era.

Inflation generically predicts a slightly red power spectrum. 
The Harrison-Zeldovich spectrum $\Delta^{2}$ in Eq. (3.21) is ruled out 
by recent observations at the $\sim 7\sigma$ confidence level [36] assuming 
the vanilla $\Lambda$CDM cosmological model and provides yet another evidence 
for the inflationary scenario. However, 
some doubts concerning the robustness of this conclusion 
have been raised in light of the recently claimed 
tension in inference of Hubble constant from cosmological 
data and local universe 
measurements , e.g. [52] and references within. 
More specifically to our model, 
taking the results of [36] at face value 
(although the standard 
cosmological model is in tension with 
local measurements) the model 
proposed here would have to be 
modified, probably via introducing a cosmological 
scale that breaks the Weyl invariance of Eq. (2.1). 
This possibility will be explored elsewhere.

In generic bouncing scenarios it is implicitly assumed 
that the bounce takes place at sufficiently large redshifts, specifically 
prior to BBN [$z_{BBN}\lesssim O(10^{10})$]. 
In that case the observed abundance 
of light elements, thermalization of the CMB, and latter processes, are explained 
exactly as in the standard cosmological model. This requirement is naturally 
satisfied by bouncing scenarios that rely on modifications of the gravitational 
field at the Planck scale which naturally take place in quantum gravity-inspired 
models. In the classical model proposed here turnaround takes place for entirely 
different reasons; $\rho_{\theta}$ is always negative and is 
characterized by $w_{\theta}=1$, i.e. $|\rho_{\theta}|$ grows 
with decreasing $\rho$ during blueshifting 
even faster than does the energy density of radiation, $\rho_{r}$. 
This implies that there is a sufficiently 
large $z_{b}>z_{BBN}$ at which $\rho_{r}(z_{b})+\rho_{\theta}(z_{b})=0$, and 
the Hubble function momentarily vanishes $\mathcal{H}(z_{b})=0$ 
while changing sign from negative to positive. Unlike 
in the quantum gravity scenarios, the requirement that $z_{b}>z_{BBN}$ sets 
a constraint on the parameter space of our model. 
In the following this constraint is derived.

From Eq. (3.21), the power spectrum of scalar metric perturbations 
generated at a time $\eta$ (during the blueshifting epoch $\eta<0$) 
is $\Delta^{2}(\eta)=\frac{9\lambda}{16\pi^{2}}
\left(\frac{|\rho_{\theta}|}{\rho_{DE}}\right)_{0}\left(\frac{\rho_{0}}{\rho(\eta)}\right)^{6}$, 
where as usual $F_{0}$ stands for the value of a function $F$ at present, $\eta_{0}$. 
Since the model is symmetric around the turnaround point 
then $\left(\frac{|\rho_{\theta}|}{\rho_{DE}}\right)_{0}$ 
stands for the ratio $\frac{|\rho_{\theta}|}{\rho_{DE}}$ at both $\eta_{0}$ and $-\eta_{0}$. 
The discussion following Eq. (3.22) implies that dynamical metric perturbations are generated only 
during the MD era. As implied in the discussion following Eq. (3.21), perturbations 
generated at $-\eta_{0}$ grow by a latter time $\eta$ 
[during which $\rho_{0}$ changed to $\rho(\eta)$] 
to $\Delta^{2}(\eta)=\frac{9\lambda}{16\pi^{2}}
\left(\frac{|\rho_{\theta}|}{\rho_{DE}}\right)_{0}\left(\frac{\rho_{0}}{\rho(\eta)}\right)^{6}$. But, this 
is also the amplitude of perturbations generated at this $\eta$. Thus, perturbations generated at 
all times in the range $-\eta_{0}<\eta<-\eta_{eq}$, where $\eta_{eq}$ is the 
time at matter-radiation 
equality, grow by the time $-\eta_{eq}$ to $\Delta^{2}(-\eta_{eq})=\frac{9\lambda}{16\pi^{2}}
\left(\frac{|\rho_{\theta}|}{\rho_{DE}}\right)_{0}\left(1+z_{eq}\right)^{6}$. 
Growth is halted during (blue- and redshifting) 
RD era. During the time periods 
$-\eta_{rec}<\eta<-\eta_{eq}$ \& $\eta_{eq}<\eta<\eta_{rec}$ the gravitational 
potential is constant as in the standard cosmological model. All this implies that the observationally 
inferred normalization of curvature perturbations 
on super-horizon scales $\Delta^{2}=2.4\times 10^{-9}$ [36], which corresponds to $z_{rec}$, 
equals $\Delta^{2}(-\eta_{eq})=\frac{9\lambda}{16\pi^{2}}
\left(\frac{|\rho_{\theta}|}{\rho_{DE}}\right)_{0}\left(1+z_{eq}\right)^{6}$. 
In the following consideration 
we make the idealized assumption that the model is 
perfectly symmetric around the turnaround point, in 
particular we assume that $\rho^{4}(\eta)\rho_{r}$ is fixed throughout 
cosmic history. Adopting $z_{eq}\sim 3400$ 
[36], and the observationally inferred $\left(\frac{\rho_{r}}{\rho_{DE}}\right)_{0}=1.3\times 10^{-4}$ 
we readily obtain $\left(\frac{|\rho_{\theta}|}{\rho_{r}}\right)_{0}=2.11\times 10^{-25}\lambda^{-1}$. 
Now, the condition $\mathcal{H}(z_{b})=0$, 
i.e. $|\rho_{\theta}|_{0}(1+z_{b})^{6}=\rho_{r,0}(1+z_{b})^{4}$ 
along with the requirement that $z_{b}>10^{10}$ implies 
that $\lambda>O(10^{-5})$. 
Therefore, the scenario proposed here is entirely consistent with the 
standard cosmological model at all epochs from BBN onwards 
insofar the dimensionless self-coupling $\lambda$ is 
sufficiently (yet reasonably) large.

Since $\varphi$ is linearly related to 
$\delta\theta$ via e.g. Eq. (3.17), and assuming all perturbations 
are linear, the former automatically 
inherits the statistical gaussian properties of the latter; 
scalar perturbations are thus expected to 
be gaussian and adiabatic (the latter property has been 
discussed in section 3.3).

\section{Summary}

While the standard cosmological model has no doubt been very successful in 
phenomenologically interpreting 
a wide spectrum of observations, it is fair to say that it still lacks a microphysical 
explanation of several key features, primarily the nature of CDM and DE.
{\it Direct} spectral information on the CMB is 
unavailable (due to opacity) in the early RD era ($z\gg 3000$). 
From the observed cosmic abundance of light elements, 
BBN at redshifts $O(10^{9})$ could be indirectly probed. 
Earlier on, at $z=O(10^{12})$ 
and $z=O(10^{15})$ [energy scales of $O(200)$ MeV 
and $O(100)$ GeV, respectively], 
the quantum chromodynamics (QCD) and electroweak phase transitions 
had presumably occurred, although their (indeed weak) 
signatures in the CMB and LSS have not been found. 
In addition, inflation, a cornerstone in the standard cosmological model, 
is clearly beyond the realm of well-established physics; its detection via 
the B-mode polarization it induces in the CMB could be achieved only 
if it took place at energy scales $\sim 13$ orders of 
magnitude larger than achievable at present. Although the inflationary 
scenario is very flexible it is also plagued with certain undesirable 
problems, such as the $\eta$-problem, trans-Planckian problem, and 
the `measure problem' in the multiverse. The latter is unavoidable 
in the currently favorite `eternal inflation' scenario. 

Ideally, an alternative 
cosmological model that agrees well with the standard cosmological 
model at BBN energies and lower, i.e. $z<10^{10}$, while 
still addressing the classical problems of the hot big bang model 
that inflation was designed to solve, and all this in the $\lesssim$ 
TeV range of energies, will definitely be an appealing alternative. 
This could be in principle achieved with a (relatively late) 
non-singular `bounce' that also removes the technically and conceptually 
undesirable initial (curvature) singularity problem of GR-based cosmological models. 
In order to achieve such a bounce 
within GR, or a conformally-related theory, certain `energy conditions' 
have to be violated. One specific realization of this program has 
been the focus of the present work.

Symmetries play a key role in our theories of fundamental interactions. 
For example, the SM of particle physics is based on a {\it local} 
$U(1)\times SU(2)\times SU(3)$ gauge group with quantizable gauge fields. 
In addition, our favorite theory of gravitation, GR, is diffeomorphism-invariant.   
In this work we entertained the possibility that in addition to 
diffeomorphism-invariance, the fundamental scalar-tensor theory 
of gravitation, Eq. (2.1), also respects a {\it global} $U(1)$ symmetry; 
the modulus of the scalar field 
is conformally coupled to gravity whereas its (dimensionless) phase is 
only minimally coupled. The cosmological model 
based on this alternative theory of gravitation has some appealing 
properties, only a few of them have been discussed in the present work. 

We have shown here that the standard cosmic scale factor 
might be replaced by the conformally
coupled modulus of the scalar field. 
Its kinetic and potential terms appear 
with the `wrong' relative 
sign in the action (thereby guarantying that gravitation is 
an attractive force), and is therefore 
inherently classical, exactly as 
the scale factor of the standard cosmological model is.
Ignoring the phase (as is {\it effectively} done in 
the standard cosmological model), which is a free quantizable 
field minimally coupled to gravitation, results in ignoring its possible 
perturbations as well, which are described by 
scale-invariant gaussian and adiabatic perturbations. 
In the standard cosmological model we 
then obtain scalar perturbations with these desired properties 
from the fluctuations of another scalar field -- the inflaton.
Perhaps even more important, the existence of this phase $\theta$ 
guarantees that cosmic history goes through 
a non-singular `bounce' rather than an initial singularity.

A tantalizing alternative scenario explored in this 
work starts with a vacuum-like- followed by MD and RD 
deflationary evolution which 
culminates at a `bounce' (essentially turnaround) 
when the (absolute value of the negative) energy 
density associated with the effective `stiff matter' 
(provided by the kinetic term of $\theta$) 
momentarily equals that of radiation. We reiterate that the various 
cosmological epochs according to this scenario only result 
from the universal evolution of masses, not space expansion -- the latter is static.
In the vacuum-like-dominated epoch the energy density of the universe is 
dominated by the quartic potential of the scalar field which is genuinely 
classical with no quantum fluctuations. 
Therefore, DE according to the present scenario is not zero-point 
energy but rather a manifestation of the self-coupling of the scalar 
field, with all other fields (e.g. Dirac, electromagnetic, etc.) playing 
a subdominant role in the dynamically evolving background at the DE era.
This DE contribution is characterized by a non-dynamical equation of state with 
no recourse to fine-tuning of the potential and with no need to introduce a 
new, e.g. quintessence, field.

During the MD blueshifting epoch, gaussian adiabatic 
scalar perturbations characterized by a flat spectrum, which are 
sourced by the (quantum) fluctuating minimally coupled field $\theta$, are generated. 
The observed power spectrum is efficiently produced during the {\it entire} 
MD deflationary epoch.
As in the case of inflation, the observed gaussianity is explained 
by the correspondence between the (quantum) 
vacuum state of an essentially free scalar field 
and the ground state of an harmonic oscillator. Adiabatic `initial' 
conditions, generically predicted by inflation, are instead 
a natural outcome of the very existence of a turnaround point in the 
present model instead of a big bang; the mere requirement of continuity 
then selects the `adiabatic' initial conditions.
Normalizing primordial scalar perturbations by their observed value 
and requiring that the turnaround takes place safely 
remote from standard BBN or any 
other lower-energy standard cosmological epochs sets a lower limit 
on the self coupling parameter of the scalar field, $\lambda\gtrsim 10^{-5}$.

The scalar field does not `slow-roll' 
along its potential (as it does in the standard inflationary scenario) 
but rather its kinetic energy is comparable to its potential 
energy at all times. It is therefore free from the fine-tuning problem 
of the inflaton potential shape generically required in the standard 
cosmological model due to radiative corrections. 
From the present work perspective `slow-roll' is an artifact of the 
standard units convention, in which all scalar fields (e.g. particle masses, 
the cosmological constant, the inflaton itself, etc.) are effectively set 
to constants. 

Linear perturbations generated during the blueshifting era 
generically survive the turnaround due to continuity of metric perturbations 
and do not undermine 
the underlying homogeneity and isotropy of the cosmological model in the 
redshifting era. 
Matter is not created in the (non-singular and adiabatic) cosmological scenario 
layed out here, nor is it destroyed. The cosmological scenario from the BBN era 
onward is exactly as in the standard cosmological 
model (assuming that $\lambda\gtrsim 10^{-5}$). 
In addition, the `anisotropy 
problem' that generally plagues bouncing scenarios does not exist in our construction.
Weyl symmetry and the consequent absence of any dimensional parameter 
in the action severely limit the possibility that such an effective 
term is present in the action. 
The existence of inverse powers of the scalar field 
at the action level is not allowed, barring the existence of any energy 
contribution characterized by $w>1/3$ in the form of perfect fluid.
In addition, although the energy density of $\theta$ behaves effectively as a perfect fluid 
with $w=1$ it does not cause any anisotropy problem simply 
because turnaround takes place 
exactly once the anisotropy-like density starts 
taking over the cosmic dynamics, and this only 
happens since this `stiff' energy density 
is negative, a consequence of the assumed $U(1)$ symmetry 
-- otherwise turnaround would not have taken place.

While conformal time is both past- and future-bounded in this scenario, 
i.e. $\eta\in(-\eta_{c},\eta_{c})$, the (effective) cosmic time in not.
There is no `horizon problem' associated with the model -- 
not for radiation, and not even for, e.g. light (but still massive) neutrinos. 
Specifically, cosmic history starts with very large 
(and in principle infinite) particle masses and therefore 
the causal horizon is much larger than would be 
naively expected from monotonically 
growing masses (that corresponds to the redshifting era), 
i.e. essentially $\eta_{0}\ll\eta_{c}+\eta_{0}$ if $\eta_{c}\gg\eta_{0}$. 
Likewise, the `flatness problem' afflicting the hot big bang scenario 
stems from the slower decay of the energy density 
associated with curvature as compared to 
that of matter in a {\it monotonically} expanding universe. 
In `bouncing' scenarios the 
situation is reversed in the pre-bounce era; 
starting at infinitely large $\rho$ 
(particle masses) one typically expects to find that the energy density in the form 
of NR matter largely exceeds that 
of curvature at any {\it finite} $\rho$ value 
in the blueshifting era. 
Since this adiabatic model is very nearly symmetric 
in $\rho$ around the turnaround point (barring entropy 
processing effects at around the RD era), 
one generally expects the universe to 
look spatially flat at any finite $\rho$ after the would-be singularity 
(actually a non-singular `bounce'). From this perspective flatness is an 
attractor-, rather than an unstable-point that requires fine-tuning. The `monopole' and 
`relic defects' problems do not arise (in the proposed scenario) 
for any reasonable $\lambda$ value (unless the latter 
is exceptionally large).

The proposed model is falsifiable in several respects: 
First, $w_{DE}=-1$ due to Weyl invariance and any observationally 
inferred $w_{DE}\neq -1$ 
would rule out the model. Likewise, CDM is made of fermionic 
particles for the same reason and thus any credible evidence in favor 
of bosonic DM would similarly rule out the model. Gaussianity 
of CMB anisotropy -- conventionally induced 
by single-field inflation models -- is yet another tenet of the 
present (effectively single-field) model. 
If non-gaussianity ultimately turns out 
to be small but non-vanishing, then it could be explained by more 
elaborate inflationary models, whereas it will 
rule out the model proposed in this work. In addition, unlike the 
inflationary paradigm that generically predicts adiabatic initial 
conditions but allows a certain level of admixture of isocurvature modes,  
the proposed `bouncing' model predicts that the initial conditions are 
purely adiabatic. Finally, within the framework adopted here, if 
B-mode polarization is ultimately 
measured, it would have to be of {\it non-quantum} origin,  
since gravitation in genuinely classical, 
the metric is unquantized, and its perturbations are not subject to the 
Bunch-Davies vacuum condition. Consequently, unlike inflationary-induced 
B-mode polarization, it does not have to be characterized by a flat spectrum.
In any case, the phase fluctuations that induce 
density perturbations in the present model do not source tensorial 
metric modes. 

We believe that, in addition to addressing the cosmological horizon, flatness 
and cosmological relic problems, the framework proposed here provides 
important insight on the nature of CDM, DE, 
initial singularity, cosmological `expansion', the flatness of the matter power 
spectrum on cosmological scales, and primordial tensor modes.
Even so, the work presented here is by no means exhaustive, and indeed 
many of its basic aspects will be further elucidated in future papers. 

\section*{Acknowledgments}
The author is indebted to Yoel Rephaeli for 
numerous constructive, critical, and 
thought-provoking discussions which were invaluable for this work.
This work has been supported by the Joan and Irwin Jacobs donor-advised fund at the
JCF (San Diego, CA).

\end{document}